\documentclass[preprint,flushrt]{aastex}
\usepackage{graphicx}
\newcommand{\eqb}{\begin{eqnarray}}
\newcommand{\eqe}{\end{eqnarray}}
\newcommand{\gesim}{\,\raisebox{-0.4ex}{$\stackrel{>}{\scriptstyle\sim}$}\,} 
\newcommand{\lesim}{\,\raisebox{-0.4ex}{$\stackrel{<}{\scriptstyle\sim}$}\,}
\newcommand{\diff}{{\rm d}}
\newcommand{\rlight}{r_{\rm L}}
\newcommand{\blight}{B_{\rm L}}
\newcommand{\brlight}{B_0}
\newcommand{\nlight}{n_{\rm L}}
\newcommand{\slight}{\sigma_{\rm L}}
\newcommand{\olight}{\omega_{\rm L}}
\newcommand{\glight}{\gamma_{\rm L}}

\newcommand{\ncold}{n'_{\rm c}}
\newcommand{\nhot}{n'_{\rm h}}

\newcommand{\rslow}{R}
\newcommand{\rstart}{r_{\rm start}}
\newcommand{\rcr}{r_{\rm cr}}
\newcommand{\rmax}{r_{\rm max}}
\newcommand{\rshock}{r_{\rm shock}}
\newcommand{\eps}{\varepsilon}

\slugcomment{{\em Publications of the Astronomical
  Society of Australia}, {\bf 18}, 415 (2001)}

\begin{document}
\title{Reconnection in pulsar winds}
\author{J. G. Kirk}
\affil{Max-Planck-Institut f\"ur Kernphysik,
Postfach 10 39 80, 69029 Heidelberg, Germany}
\email{John.Kirk@mpi-hd.mpg.de}
\and
\author{Y. Lyubarsky} 
\affil{Department of Physics, Ben-Gurion University, P.O. Box 653, Beer Sheva 84105, Israel}
\email{lyub@bgumail.bgu.ac.il}
\begin{abstract}
The spin-down power of a pulsar 
is thought to be carried away in an MHD wind in which, 
at least close to the star, the 
energy transport is 
dominated by Poynting flux. 
The pulsar drives a low-frequency wave in this wind, 
consisting of stripes of 
toroidal magnetic field of alternating polarity, propagating in a 
region around the equatorial plane. 
The current implied by this configuration falls off more slowly with radius 
than the number of charged particles available to carry it, so that the 
MHD picture must, at some point, fail.
Recently, 
magnetic reconnection 
in such a structure has been shown to 
accelerate the wind significantly. 
This reduces the magnetic field in the comoving frame and, consequently, the 
required current, enabling the solution to
extend to much larger radius.
This scenario is discussed and, for the Crab Nebula, the range of 
validity of the MHD solution is compared with the radius 
at which the flow appears to terminate. For sufficiently high particle
densities, it is shown that a low frequency entropy wave can propagate 
out to the termination point. In this case, the \lq termination shock\rq\ 
itself must be responsible for dissipating the wave.\\
{\em Dedicated to Don Melrose on his 60th birthday}
\end{abstract}
\keywords{pulsars: general---pulsars: individual (Crab)---MHD---stars: winds
  and outflows---plasmas---waves}

\section{Introduction}
\label{introduction}
The supply of relativistic electrons and magnetic field needed to provide the 
synchrotron radiation observed from the Crab Nebula has long been thought to
originate in a central star (Piddington~1957; Kardashev~1964). 
With the identification of this
star as a pulsar, the earlier suggestion (Pacini~1967)
that the star also 
emits a wave, whose energy is presumably released into the Nebula, 
attracted considerable attention (Ostriker \& Gunn~1969; Rees \& Gunn 1974).
As an energy source for its environment, the Crab pulsar is by no means unique.
The pulses of electromagnetic radiation emitted by 
other pulsars contain 
only a fraction of the power which the neutron star is apparently
releasing from its store of rotational energy. 
In several cases, nebular emission is observed 
to surround the pulsar, but, even where it is not, 
it is generally thought that most of the spin-down power is
carried away from the star by a relativistic wind consisting of a mixture
of particles, waves and magnetic field.
Attempts to provide a consistent description of this wind touch upon a 
fundamental problem in electrodynamics --- the dichotomy between a single
particle approach and a continuum description. 
In the pulsar wind case, one can either
start from the solution describing a rotating magnetised object, compute 
individual particle trajectories and treat their 
combined effect as a perturbation, or, alternatively,
consider the particles as a fluid and construct, for example, 
a solution with 
an ideal
MHD wind emerging from a magnetised rotating star.

In the former case, it has been found that the wave component of the vacuum 
fields transfer energy
rapidly to the particles. The resulting damping is strong and the wave
will not propagate if the particle density exceeds a critical value
(Asseo et al~1978; Leboeuf et al~1982; Melatos \& Melrose~1996). 
Since the particle density decreases as the radius increases, the 
vacuum wave-like solution exists only outside a critical radius.
In the fluid case, one faces the difficulty of formulating a generalised Ohm's
law. The simplest choice of infinite conductivity 
corresponds to ideal MHD, but has the problem that 
it may imply currents too large to be carried by the finite number of
charged particles in the fluid. Given the structure of the magnetic field, 
the current densities can be computed 
from Amp\`ere's law. These generally decrease with radius less rapidly
than the decrease in the density of charged particles available to carry the
current (Usov~1975; Michel~1982). 
Since the velocity of the current carriers cannot exceed that of 
light, the MHD approach is valid only inside a critical radius.
Thus, there are theoretical methods to describe
both the inner and the outer parts 
of a pulsar driven wind. However, as yet, no way has been found of
linking these two regimes.

In the case of the Crab Nebula, observations suggest that
the wind energy is dissipated into relativistic particles 
at a radius $r\approx10^9\rlight$, where
$\rlight=c/\Omega$ is the radius of the light cylinder and $\Omega$ the
angular velocity of the neutron star. 
The physics of this region
(which we will term \lq termination shock\rq) is
clearly an important problem. As a preliminary step, 
it is of interest to examine whether
the wind solutions enable one to make a statement about the 
properties of the flow immediately before entering the shock region.
This is the problem we address in this paper. We discuss 
a highly simplified non-axisymmetric ideal MHD wind --- a \lq
striped wind\rq\ --- and describe how a model of magnetic reconnection 
(Coroniti~1990; Michel~1994) modifies the wind dynamics 
(Lyubarsky \& Kirk~2001, henceforth LK). 
As a result, the range of validity of the MHD
solution is extended. 
Applying the results to the Crab pulsar wind, we find that
for sufficiently high particle densities the solution is valid up to  
the location of the termination shock.
We then speculate that the crucial physical
process at work in the shock is the dissipation of the 
wave energy carried by the MHD wind.

\section{The MHD wind}
\label{coroniti}
An exact analytic
solution for an MHD pulsar wind has been found
only for the idealised case of a monopole magnetic field 
in the force-free limit (Michel 1973). This solution has a very simple 
structure: the flow does not collimate, i.e., both the velocity and the
magnetic field have purely radial components in the
meridional plane, $B_\theta=0$, $v_\theta=0$ and there exist no closed
field lines. A more realistic case would allow for closed field lines
near the star, but, outside this region, it might be expected that the flow
along open field lines mimics the monopole case, perhaps with a modified
dependence on latitude. The obvious inconsistency of
using a magnetic monopole as a source can be lifted by 
introducing the \lq split monopole\rq. In this case 
the system retains axial symmetry; 
the sign
of each magnetic field component changes 
at the equator of the star, all other quantities
remaining the same. A current sheet is then implied, which 
lies in the equatorial plane and 
separates the oppositely directed toroidal field
components in the northern and southern hemispheres.
It is also possible to lift the force-free approximation, replacing 
it by an ultra-relativistic
approximation
(Bogovalov 1997) to find that 
highly super-Alfv\'enic, relativistic winds also 
collimate only very slightly
and the radial velocity 
pattern remains a good
approximation, even when particle
inertia is taken into account (Beskin et al~1998; 
Chiueh et al~1998; Bogovalov
\& Tsinganos~1999). 

In a radial wind,
the outward pressure gradient exerted by the magnetic field is precisely
compensated by the inward tension force, so that the wind speed remains
constant, as does the quantity $\sigma$, defined as the ratio of the Poynting
flux to the energy flux in particles.
Thus, a steady
MHD wind driven by a magnetised neutron star carries energy 
predominantly in the form of
Poynting flux, which is not converted into kinetic energy
anywhere in the flow.   
This is a major difficulty, because modelling of the synchrotron emission
of the Crab Nebula (Rees \& Gunn~1974; Kennel \& Coroniti~1984; 
Emmering \& Chevalier~1987) suggests
that if the wind terminates at a relativistic MHD shock, the 
energy flux must be carried predominantly by particles. 

\begin{figure}
\begin{center}
\includegraphics[bb=143 303 488 517,width=.7\textwidth]{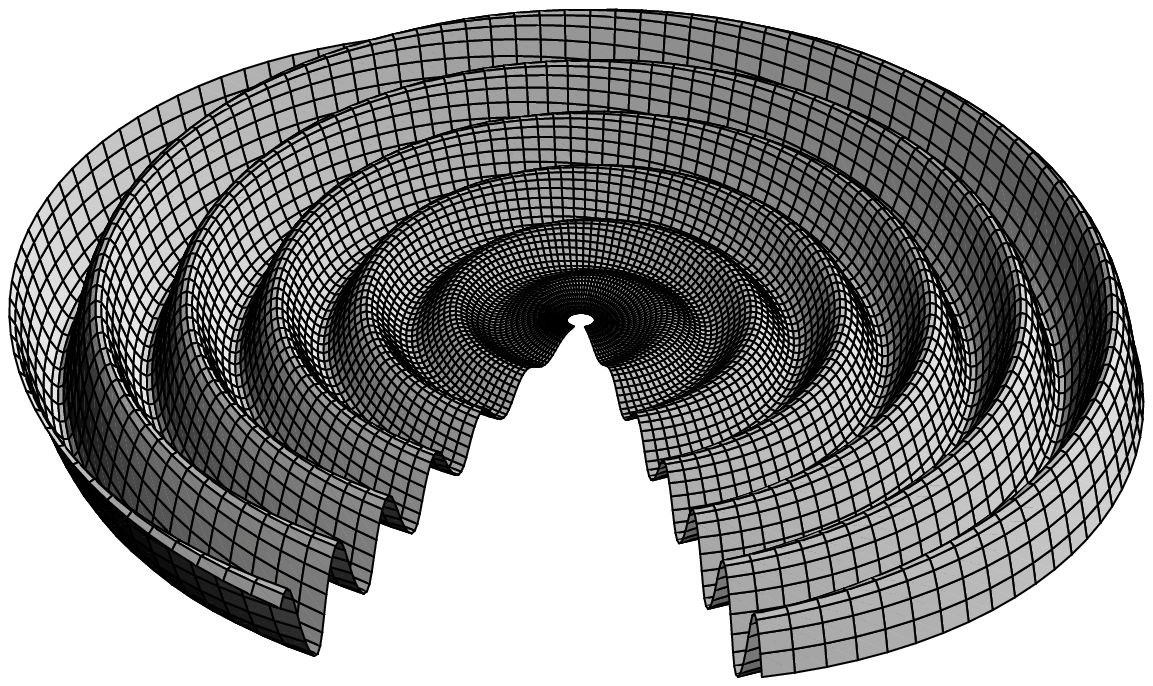}\\
\includegraphics[bb=143 271 488 550,width=.7\textwidth]{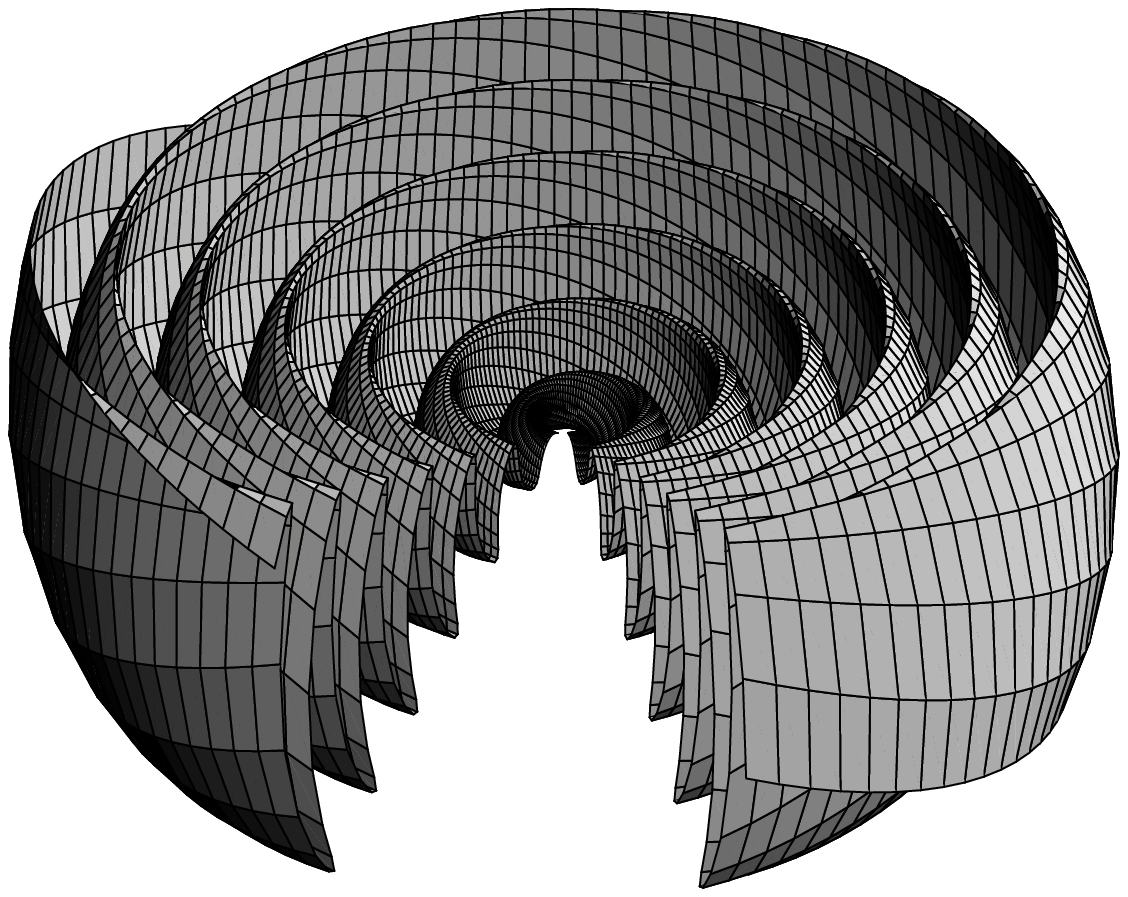}
\end{center}
\caption{\protect\label{sheet}
The current sheet in the split-monopole solution for a 
radial relativistic wind. The 
sheet has been truncated at a distance from the axis of $40\rlight$, and a
sector has been removed for clarity. The
wavelength of the corrugation is $2\pi\rlight$.
The upper figure displays the sheet for an
inclination angle $\chi$ 
between the rotation and magnetic axes of $6^{\rm o}$, the 
lower for an inclination angle of $26^{\rm o}$.}
\end{figure}

The rotating neutron star of a pulsar does not have an axisymmetric
magnetic field, so that an axisymmetric wind is quite possibly a poor
approximation. 
Non-axisymmetric solutions 
are, of course, very much more 
difficult to find. Fortunately, however, the technique 
of replacing a monopole source 
by a split monopole can also be used
when the magnetic axis does not lie along the rotation axis 
(Bogovalov~1999). The solution is 
essentially the same as in the monopole case: the velocity field is unchanged,
the pattern of the magnetic field lines is the same, but the direction 
(sign) of the field depends on the hemisphere in which the field line has 
its anchor point on the star. The current sheet, which lies in the equatorial
plane in the axisymmetric case, takes on the form of an outward moving, 
spiral corrugation, as shown in Fig.~\ref{sheet}.

In the ultra-relativistic, highly super-Alfv\'enic limit, the solution 
in terms of spherical polar coordinates $(r,\theta,\phi)$
is given by:
\eqb
\begin{array}{lll}
u_r=\beta\glight&u_\theta=0&u_\phi=0\\
B_r=\pm \brlight(\rlight/r)^2&B_\theta=0&B_\phi=\mp(\rlight\sin\theta/\beta r)\brlight\\
E_r=0&E_\theta=\mp\brlight\sin\theta(\rlight/r)&E_\phi=0
\end{array}
\label{urlimit}
\eqe
(Bogovalov~1999), where $u_{r,\theta,\phi}$ are components of the four
velocity, $B$ and $E$ are the magnetic and electric fields and $\brlight$
is the magnitude of the radial component of ${\bf B}$  at the light cylinder, 
$r=\rlight$.
The signs of $B_r$, $B_\phi$ and $E_\theta$ depend on whether the field line
lies above (upper sign) or below (lower sign) 
the current sheet. Associated with this
solution are distributed (\lq body\rq) currents and
charges as well as surface currents and charges located in the sheet. 
The body current is 
found from the curl of ${\bf B}$,
(since the displacement current vanishes outside the sheet) and the body
charge follows from Poisson's equation:
\eqb
\begin{array}{lll}
j_r=-\left(c\brlight\cos\theta/2\pi\beta\rlight\right)\left(\rlight/
    r\right)^2&
j_\theta=0&j_\phi=0\\
\rho=\beta j_r/c
\end{array}
\eqe
The current sheet is located at the position $\Phi({\bf r})=0$, with 
\eqb
\Phi&=&\cos\chi\cos\theta+
\sin\chi\sin\theta\cos\left({r-\beta ct\over\beta\rlight}+\phi\right)
\label{sheeteq}
\eqe
where $\chi$ is the angle between the magnetic symmetry axis and the rotation
axis and $\beta c\approx c$ is the 3-velocity of the wind. 
The surface current carried in the sheet is directed perpendicular to the 
adjacent magnetic field, and has the magnitude
\eqb
k={c[1-\beta^2({\bf \hat n}\cdot{\bf \hat r})^2 ]\over 2\pi}
\left|{\bf \hat n}\times {\bf B}\right|
\label{surfacecurrent}
\eqe
where ${\bf \hat n}={\bf \nabla}\Phi
/|{\bf \nabla}\Phi|$ is a unit vector normal to the sheet
[e.g., Jackson (1975), pp 20--22].
Except at points where the sheet normal is perpendicular to ${\bf \hat r}$,
the surface current scales with $1/r$, whereas the body current scales
according to $j_r\propto 1/r^2$. The particle density in the wind drops
off as $1/r^{2}$,  so that the charge carriers 
can maintain the body currents without
an increase in velocity. To maintain the surface currents, however, 
an increase with radius of either the surface density of charge 
carriers, or of their velocity is required. The MHD solution 
fails if 
equation (\ref{surfacecurrent}) requires a current too large to be carried
even if the particles contained in the sheet 
move at the speed of light. 

\section{Reconnection}
\begin{figure}
\begin{center}
\includegraphics[bb=188 396 522 625,width=.9\textwidth]{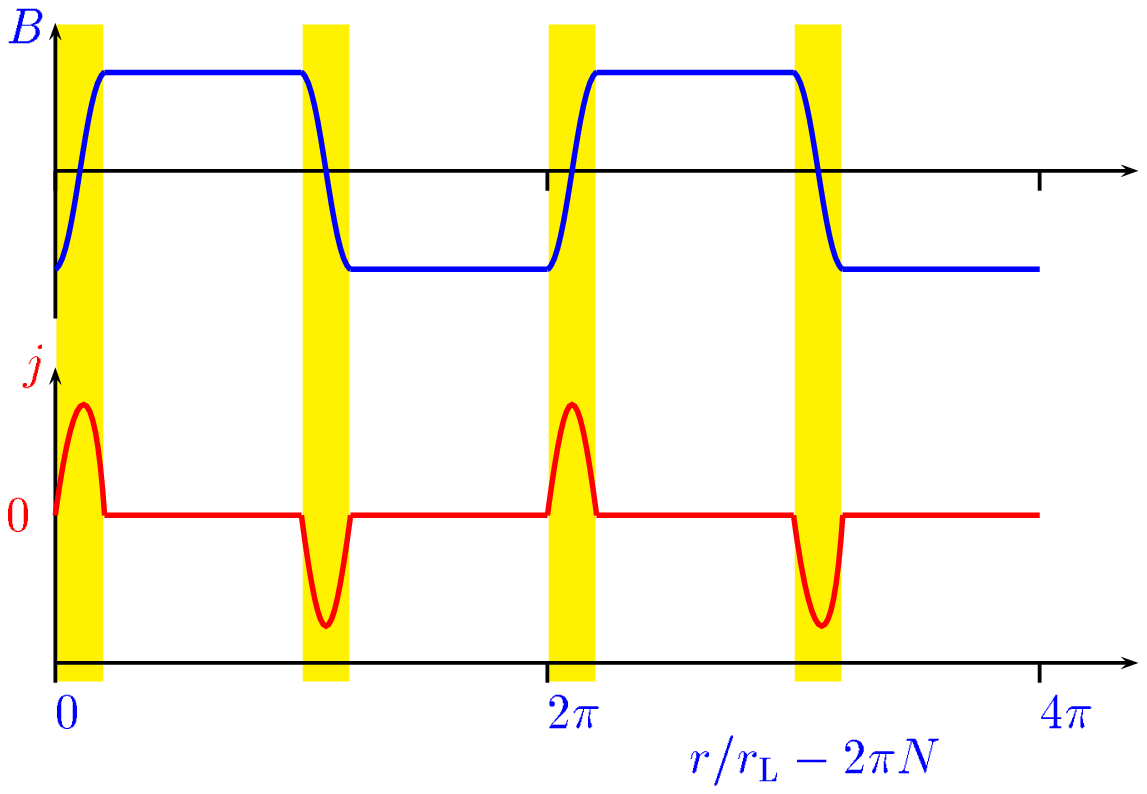}\\
\end{center}
\caption{\protect\label{sketch2}
A schematic sketch of the 
radial structure of the magnetic field and current density in the wind 
far outside the light-cylinder ($N\gg1$). In the equatorial plane, the 
current sheets are equally spaced, as shown. However, except in the singular
case $\chi=90^{\rm o}$, the intersheet spacing
depends on the magnetic polarity away from the equator and the sheets vanish
altogether for $|90^{\rm o}-\theta|> \chi$}.
\end{figure}

At large radius, the corrugated current sheet described above 
resembles locally, 
a set of concentric nested spheres, when viewed at latitudes close 
to the equator ($|90^{\rm o}-\theta|\ll\chi$).  
The dynamics of the sheet may
then be considered using a one-dimensional (radial) treatment.
In the
equatorial plane, the sheets of opposite polarity are evenly spaced. A sketch
of the configuration at $r\gg\rlight$ is shown in Fig.~\ref{sketch2}. 

According to Amp\`ere's law, the integral of the 
current density across a current sheet is proportional to the 
change in magnetic field strength, measured in the comoving frame.
In a radial wind, the toroidal field drops as $1/r$, and the particle density
as $1/r^2$, when both quantities are measured in the lab.\ frame. Thus,
in a wind of constant speed, the charge carriers in the sheet are forced to  
counter-stream with higher and higher velocities as the plasma moves outwards.
This situation is likely to lead to instability, which could, in turn,
provide an anomalous resistivity and hence magnetic reconnection.
Coroniti (1990) [see also Michel (1994)] proposed a simple picture intended to 
capture the physics
of magnetic reconnection in the sheets. The suggestion is that the sheets
attain a minimum thickness equal to the gyro-radius which a particle in
the hottest part of the sheet would have, if it were moving in the magnetic
field adjacent to the sheets. Equivalently (to within a factor of order
unity), the thickness can be set by 
demanding that the maximum current density in the sheets is 
equal to the product of the particle density, the electronic charge and the
velocity of light. 
This enables one to formulate a system of equations for a  
wind that 
consists of two phases: a hot unmagnetised phase (corresponding to the
current sheets) and a cold magnetised phase. 
From Fig.~\ref{sheet}, it can be seen that, for an oblique rotator, the points
at which a radius vector cuts the current sheet are equally spaced only in the
equatorial plane. At other latitudes, the cold phase of one polarity
dominates. In order to maintain pressure equilibrium, 
the magnetic field strength in the cold phase must remain constant on
the scale of a wavelength, so that the cold phase of 
the dominant polarity is {\em thicker} than that of the opposite polarity.
At a critical latitude ($\theta=90^{\rm o}-\chi$) the radius vector grazes the
current sheet, and for $|90^{\rm o}-\theta|>\chi$ the hot phase is absent.

In Coroniti's original work, 
the density in the hot and cold phases was assumed equal. However, this leads
to inconsistencies and an incorrect picture of the evolution of the wind.
Recently, Lyubarsky \& Kirk (LK) have used a short wavelength approximation
to analyse the evolution of the pattern of hot and cold phases, which is
simply an entropy wave in the wind plasma. 
Reconnection at the sheet boundaries 
starts at a certain radius $r=\rstart$, which depends on how many 
particles are contained in the current sheet initially.
The short wavelength approximation consists in assuming this 
radius to lie well 
outside the light-cylinder: $\eps\equiv \rlight/\rstart\ll 1$. 
The slow evolution
of the wave is governed by a set of five equations. Three of these 
are conveniently written in differential form. These are the 
conservation of
particle number and of energy,
\eqb
{\partial\over\partial\rslow}
\left\lbrace
   \rslow^2\gamma_0v_0
   \left[
      \left(
      1-\Delta
      \right)
   \ncold+\Delta\nhot
   \right]
\right\rbrace&=&0
\label{continuity}\\
{\partial\over\partial\rslow}
\left\lbrace
\rslow^2\gamma_0^2v_0 mc^2
\left[\left(1-\Delta\right)\ncold+\Delta\nhot\right]
+2\rslow^2\gamma_0^2v_0\left(1+\Delta\right){B'^2\over8\pi} \right\rbrace&=&0
\label{energy}
\eqe
and the entropy equation
\eqb
{4\Delta B'^2\over \rslow^2}{\partial\over\partial\rslow}
\left(\rslow^2 \gamma_0v_0\right)
+3\gamma_0v_0{\partial\over\partial\rslow}\left(\Delta B'^2\right)
+&&
\nonumber\\
{v_0\over\gamma_0}{\partial\over\partial\rslow}
\left[\gamma_0^2\left(1-\Delta\right)B'^2\right]
+{2\gamma_0 B'^2 
\left(1-\Delta\right)\over\rslow}{\partial\over\partial\rslow}
\left(\rslow v_0\right)
&=&0
\enspace.
\label{entropy}
\eqe
Here, $\gamma_0(R)$ and $v_0(R)$ are the zeroth order (in terms of $\eps$) 
Lorentz factor and (3--)velocity of the 
flow, the dimensionless radius variable is $\rslow=r/\rstart$,
$\Delta(R)$ is the fraction of a wavelength occupied by the sheets, 
$\ncold(R)$ and $\nhot(R)$ are the (zeroth order) 
proper number densities in the cold 
and hot phases, respectively, and $B(R)=\gamma_0 B'(R)$ is the 
(zeroth order) magnetic field in the 
cold phase (the prime indicates a quantity measured in the plasma 
rest frame). 
The condition 
of ideal MHD (flux freezing) in the cold phase:
\eqb
{B\over r n_{\rm c}}&=&{\rm constant}
\label{freeze}
\eqe
follows from the zeroth order equation of continuity, 
$r^2v_0\ncold={\rm constant}$ and the zeroth order Faraday equation,
${1\over r}{\partial\over\partial r} rE=-{1\over c}{\partial \over
\partial t}B=0$
together with the ideal MHD condition $E=-vB$.
The remaining equation stems from 
the requirement that the sheet thickness equal the gyro radius:
\eqb
\Delta&=&{r\ncold c\over 4\pi\nhot\gamma_0 v_0\rlight\kappa}
\label{coroniticond}
\eqe
In this equation, we have introduced the multiplicity parameter $\kappa$, 
which is a dimensionless measure of the number density in the 
cold phase of the wind, in terms of the Goldreich-Julian density (see LK):
\eqb
\kappa&=&{e c \nlight\over\blight P}
\eqe
where $P$ is the pulsar period, $\nlight=n_i/\eps^2$ and $\blight=B_i/\eps$
are the density (in the cold phase) and magnetic field, extrapolated back to
the light cylinder, in terms of the values $n_i$, $B_i$ at $r=\rstart$.

The set of differential 
equations (\ref{continuity}), (\ref{energy}) and (\ref{entropy}) 
together with 
(\ref{freeze}) and (\ref{coroniticond}) have been solved by
LK. The solution is specified by three parameters, for
example, 
the Lorentz factor $\glight$, the multiplicity $\kappa$ and the ratio of the
particle gyro frequency $\olight$ at $r=\rlight$
to the rotation frequency of the pulsar $\olight/\Omega$. 
The values of the first two parameters are uncertain, but $\olight/\Omega$
is more or less directly accessible from observation and is, 
to within a factor of a few, equal to the potential difference 
generated across open field lines, in units of the electron rest mass.
(For the Crab pulsar $\olight/\Omega\approx 10^{11}$.)
As an alternative to the multiplicity, one may use
the magnetisation parameter $\slight$,
defined as the ratio at the light-cylinder
of Poynting flux to particle energy flux in the cold
phase of the wind, and related to the above parameters by
$\slight=\olight/(2 \kappa\glight\Omega)$. 
In addition, the initial condition $r_{\rm rstart}$ must be given. 

\begin{figure}
\begin{center}
\includegraphics[bb=27 155 580 708,width=.9\textwidth]{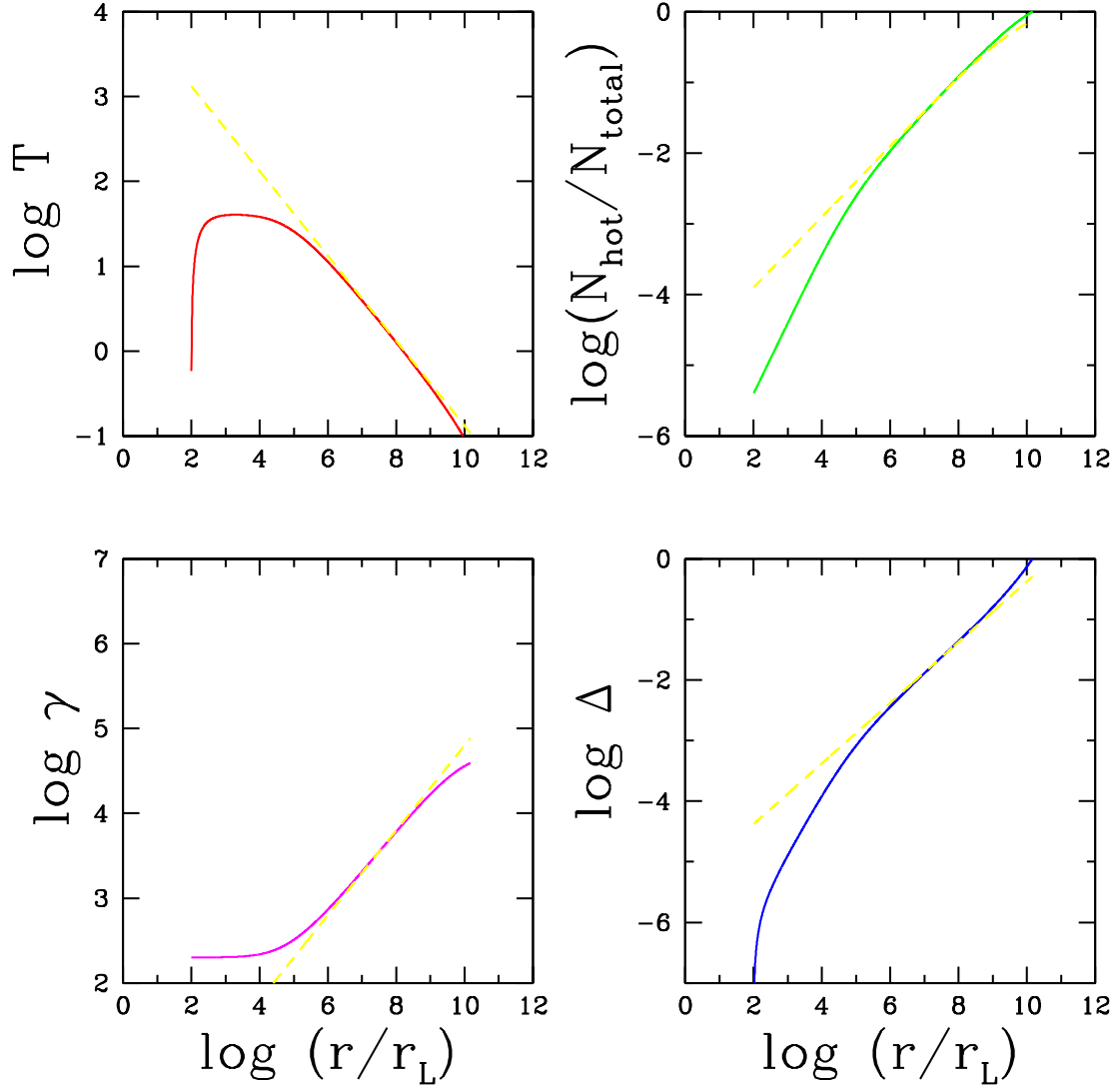}\\
\end{center}
\caption{\protect\label{numsol}
A numerical solution of the reconnecting striped wind problem, using
parameters appropriate for the Crab pulsar:
$\olight/\Omega=10^{11}$, $\kappa=10^5$, $\glight=100$, corresponding to
$\slight=5000$, and assuming reconnection starts at $\rstart=100\rlight$. The dashed
line shows the asymptotic solution of LK.
}
\end{figure}

In Fig.~\ref{numsol} we display an example of a solution appropriate for a
pulsar that generates a maximum potential equal to that of the Crab pulsar:
$\olight/\Omega=10^{11}$. The multiplicity is chosen
to be $\kappa=10^5$, and the initial Lorentz factor is $\glight=100$, 
(corresponding to $\slight=5000$). Reconnection is assumed to commence 
at $\rstart=100\rlight$, i.e., $\eps=0.01$. 
As the magnetic flux is dissipated, the hot phase expands and performs work on
the wind, causing it to accelerate. 
The system quickly
approaches the asymptotic solution given by LK and shown
as dashed lines. This solution reads:
\eqb
\gamma&=&{1\over \kappa}\sqrt{\olight r\over8\pi\Omega\rlight}
\label{gammaasympt}
\\
\Delta&=&\sqrt{r\Omega\over18\pi\rlight\olight}
\label{deltaasympt}\\
{T'\over mc^2}&=&\sqrt{\pi\olight\rlight\over18\Omega r}
\label{tempasympt}\\
p'&=&8\pi\left({\blight^2\over 8\pi}\right)
\left({r\over\rlight}\right)^{-3}\kappa^2
\left({\olight\over\Omega}\right)^{-1}
\label{pressasympt}
\eqe
where $p'=B'^2/8\pi$ is the pressure of the hot phase and $T'$ its
temperature, in energy units.

\section{Validity of the solution}

The ideal MHD part of the solution, in which the radial velocity is constant
and no magnetic flux is dissipated, is valid, by assumption, for 
$r<\rstart$. This range is determined by the surface density of charges
assumed to be present in the quasi-spherical current sheets when they are
first established. For $r<\rstart$ 
the current sheet contains a sufficient 
number of particles to supply the required current.
Provided the initial surface density is small compared to the surface
density of particles in the cold, magnetised phase (i.e., the integral of
$n_{\rm h}$ over one wavelength) the sheets are able to maintain the required
current at $r>\rstart$ by expanding and absorbing particles from the cold
phase. In this way, the MHD picture of the wind retains validity,
although the flow necessarily contains non-ideal regions in which oppositely
directed magnetic flux is steadily annihilated. The solutions can be followed 
until the current sheets --- as seen in the one-dimensional description --- 
appear to merge. In the equatorial plane, this occurs 
at a radius given approximately by
\eqb
\rmax&\approx&{\olight\over\Omega}\rlight
\label{rmax}
\eqe
and implies the
complete annihilation of the magnetic field. Above and below the equatorial
plane,
the current sheets are not equally spaced, and the total magnetic flux passing
through one complete wavelength of the pattern is non-zero. Since this flux is
conserved in our picture, reconnection leaves behind a
residual magnetic field, whose magnitude 
can be found from Eq.~(\ref{sheeteq}) and is  
\eqb
B_{\rm res}&=& {\rlight\blight\over r}\left(
{2\arccos(\cot\chi\cot\theta)\over\pi} - 1\right)
\eqe
Beyond $\rmax$, 
the \lq current sheet\rq\  
is no longer corrugated, but lies in the
equatorial plane. Since the toroidal magnetic field goes smoothly to zero 
as this plane is approached, the current it carries is proportional to the 
radial component of the field, which varies as $r^{-2}$.
As in the axisymmetric, split-monopole case, the 
MHD picture is still valid, since the number of available particles is always
sufficient to provide this current.

However, there are two limitations which restrict the validity of this
solution at large radius. The first is the assumption, implicit in
Eq.~(\ref{coroniticond}), that the electrons in the hot phase are
relativistic. This is a relatively minor technical limitation of the 
treatment, which is encountered for radii more than a few percent of 
$\rmax$. The second limitation is more subtle, and can be stated in a
number of ways. It also arises from Eq.~(\ref{coroniticond}). In  
prescribing 
the thickness of the sheet in terms of the densities in the 
two phases and the wind velocity, 
the movement of the sheet edge is decoupled from the local wave speeds, and 
can even move superluminally. Clearly, a physically realistic
prescription would involve, at the very least, 
an additional dissipation of energy into waves
when the velocity of the sheet edge approaches $c$.   
This point is reached when
\eqb
v_{\rm edge}&=& v_0\left(1 + {\pi\rlight\over 2}
{\diff \Delta\over\diff r}\right)\,=\,c
\eqe
which can be written, using Eq.~(\ref{deltaasympt})
\eqb
\Delta&=&{2 r\over \pi\gamma_0^2\rlight}
\eqe
To within a factor of the order of unity, this is coincident with the 
limit discussed in LK [Eq. (36)]. Using the asymptotic
solution, the critical distance $\rcr$ beyond which the solution 
fails is
\eqb
\rcr&=&1.5\times 10^3 \kappa^4\rlight \left({\olight\over\Omega}\right)^{-1}
\label{rcr}
\eqe
Combining this with equation~(\ref{rmax}), we find
\eqb
{\rcr\over\rmax}&=&1.5\times
10^3\kappa^4\left({\olight\over\Omega}\right)^{-2}
\label{kappacond}
\eqe

\section{Discussion}

The inclusion of non-ideal MHD effects --- albeit phenomenologically ---
results in a substantial extension beyond $\rstart$ 
of the range of validity of the \lq inner\rq\ wind solution, which
can be described using a continuum picture. Starting with an entropy wave
outside the light cylinder, this solution
can exist up to a maximum radius $\sim\rmax$ (Eq.~\ref{rmax}) 
and convert a substantial part of the total 
energy flux into bulk kinetic energy of the particles, provided 
$\kappa\gesim0.1\sqrt{\olight/\Omega}$
[see eq.~(\ref{kappacond})].
The situation
in the outer parts of the wind then resembles the axisymmetric, split
monopole solution, except that the magnitude of the toroidal component of the
magnetic field passes smoothly through zero on crossing the 
equatorial plane.
However, if $\kappa\lesim0.1\sqrt{\olight/\Omega}$, the solution loses
validity at
$r\sim\rcr$ (Eq.~\ref{rcr}) and the energy 
contained in the non-axisymmetric or (entropy) wave
component must be channelled into another type of wave and/or 
into the particles. 

In reality, the wind driven by a pulsar 
is confined by the surrounding medium, which
may disrupt the solution either before reconnection is complete or before 
the MHD approach loses validity.
Usually, disruption is envisaged to occur at a \lq termination shock\rq\ where
the flow makes an abrupt transition from supersonic 
to subsonic velocity. In the 
case of the Crab Nebula, this transition can be identified with a sharp
increase in the synchrotron emissivity at a radius of approximately
$\rshock\equiv10^9\rlight$. 
The Crab pulsar has $\olight/\Omega=10^{11}$, so that, 
according to Eq.~(\ref{rmax}), $\rmax\gg\rshock$ and 
very little of the striped magnetic flux has time to reconnect.
For the MHD solution to remain valid up to  
$\rshock$ a minimum number of particles is required. 
From Eq.~(\ref{rcr}) this may be expressed as a condition on the multiplicity
parameter: $\kappa\gesim1.6\times10^4$. 
Assuming that this condition is satisfied, 
the region $r<\rshock$ can be described by the striped MHD wind. Close to the
equator, most of the energy flux is carried by the magnetic field of this
entropy wave. If, in this region, 
the termination shock is to dissipate a substantial
fraction of the wind energy, it must not only decelerate the wind to subsonic
velocity, but also damp the oscillating component of the magnetic field.
This conclusion is consistent with the constraints from modelling the
synchrotron emission of the nebula, which indicate that the energy density 
of the plasma downstream of the shock is not dominated by the magnetic field.
The physics of such a transition is at present unclear, but is 
likely to have profound implications for particle acceleration. 

\acknowledgements{We are grateful to L. Ball, Y. Gallant, J. Kuijpers, 
A. Melatos and D. Melrose for helpful discussions.}

\end{document}